\documentclass[12pt,preprint]{aastex}

\shorttitle{r- and s-process Elemental Abundances}
\shortauthors{Bond et al.}

\begin{document}

\title{Beyond the Iron Peak: r- and s-process Elemental Abundances in Stars with Planets}

\author{J. C. Bond\altaffilmark{1}, D. S. Lauretta\altaffilmark{1}, C. G. Tinney\altaffilmark{2,3}, R. P. Butler\altaffilmark{4}, G. W. Marcy\altaffilmark{5}, H. R. A. Jones\altaffilmark{6}, B. D. Carter\altaffilmark{7}, S. J. O'Toole\altaffilmark{3} $\&$ J.
Bailey\altaffilmark{8}}
\altaffiltext{1}{Lunar \& Planetary Laboratory, University of
Arizona, 1629 E. University Blvd, Tucson, AZ 85721-0092}
\altaffiltext{2}{Dept. of Astrophysics, University of NSW, NSW
2052, Australia} \altaffiltext{3}{Anglo-Australian Observatory, PO
Box 296, Epping, NSW 1710, Australia} \altaffiltext{4}{Department
of Terrestrial Magnetism, Carnegie Institution of Washington, 5241
Broad Branch Road NW, Washington, DC 20015-1305}
\altaffiltext{5}{Department of Astronomy, University of
California, Berkeley, CA 94720} \altaffiltext{6}{Centre for
Astrophysics Research, University of Hertfordshire, College Lane,
Hatfield AL10 9AB, England} \altaffiltext{7}{Faculty of Sciences,
University of Southern Queensland, Toowoomba 4350, Australia}
\altaffiltext{8}{Physics Department, Macquarie
University, Sydney, NSW 2109, Australia}

\begin{abstract}
We present elemental abundances of 118 stars (28 of which are
known extrasolar planetary host stars) observed as part of the
Anglo-Australian Planet Search. Abundances of O, Mg, Cr, Y, Zr,
Ba, Nd and Eu (along with previously published abundances for C
and Si) are presented. This study is one of the first to
specifically examine planetary host stars for the heavy elements
produced by neutron capture reactions. We find that the host stars
are chemically different to both the standard solar abundance and
non-host stars in all elements studied, with enrichments over
non-host stars ranging from 0.06 dex (for O) to 0.11 dex (for Cr and Y).
Such abundance trends are in agreement with other previous studies
of field stars and lead us to conclude that the chemical anomalies
observed in planetary host stars are the result of normal galactic
chemical evolution processes. Based on this observation, we
conclude that the observed chemical traits of planetary host stars
are primordial in origin, coming from the original nebula and not
from a ``pollution'' process occurring during or after formation
and that planet formation occurs naturally with the evolution of
stellar material.
\end{abstract}


\section{Introduction}
Extrasolar planets are known to preferentially orbit
metal-enhanced stars. Chemical analyses of known host stars have
shown that they appear to be metal enriched compared to a sample
of ``average'' F, G and K stars not known to harbour planets
\citep{g1,g2,g4,g5,g3,sb,s1,s04,g6,sm,re,fv,bond}. In addition to
this global metal enrichment, individual elements have also been
shown to exhibit similar trends, although not as statistically
significant or with such large host and non-host differences
(\citealt{g6,sb,bo,fv,bond}).

The vast majority of abundance studies completed so far have
focussed on elements  with atomic number (Z)$\leq$30 (i.e. those
located before the iron stability peak). These elements are
produced by a variety of processes (for example alpha particle
addition, the CNO cycle and stellar burning reactions) in stellar
interiors during main sequence evolution. Elements located beyond
the iron peak, however, are produced via neutron-capture
reactions, specifically the rapid (r-) and slow (s-) processes.
Here rapid and slow refers to the speed of neutron capture with
respect to the $\beta$-decay rate of the nuclei. In the r-process,
neutron capture occurs before $\beta$-decay of the unstable nuclei
can occur. Alternatively, in the s-process neutron capture occurs
less frequently, thus allowing the nuclei to undergo $\beta$-decay
before capturing another neutron and effectively allowing the
nuclide to remain within the valley of $\beta$ stability.

Due to the different neutron fluxes required for each process
($\sim$10$^{23}$ neutrons\,cm$^{-2}$\,s$^{-1}$ for the r-process
vs. $\sim$10$^5$ neutrons\,cm$^{-2}$\,s$^{-1}$ for the s-process
(\citealt{clay})), the r- and s-processes occur in different
stellar environments. There is some debate as to exactly where the
r-process occurs (see e.g. \citealt{qian}), but it is believed to
primarily occur in type II supernova events. The s-process is
thought to produce its heavier elements (such as Ba and Ce) in the
interior of lower mass AGB stars and its lighter elements (such as
Sr, Y and Zr) during the He-burning stages of stellar evolution
for larger mass stars (\citealt{red03}). Due to these vastly
different settings, the abundances of the heavy elements produced
by these processes can provide information on the history of the
material later incorporated into both the host star and planets
themselves, as well as testing models of galactic chemical
evolution (e.g. \citealt{fen} and \citealt{lan}). These elements
have been part of several previous spectroscopic studies of stars
without planets (e.g. \citealt{ed, red03,ap,ben,red06}), however
only Ba (\citealt{hu}) and Eu (\citealt{gonz}) have been
specifically examined in a small number of known extrasolar
planetary host stars.

Studies of stellar abundances have become critical to our
understanding of planetary formation processes. However, in spite
of significant advances in atmospheric models, stellar interiors
and atomic line data in recent years, the measurement of stellar
abundances is still an intricate process. In particular, choices
made in the way an analysis is carried out can result in
systematic abundance differences on sizes similar to the effects
we would most like to understand in the stars themselves.
Additionally, in many cases there is no consensus on the
``correct'' choice in these techniques.

To make headway in this field, therefore, it is essential to
perform abundance studies in a manner immune to such systematic
errors, by analysing both samples of interest, and control
samples, in an identical manner. For the abundances of
exoplanetary host stars, that means both host and non-host stars
must be analysed in an identical manner. This is the primary goal
of this current study.

We have derived abundances of five r-and s-process elements, in
addition to three other lighter elements, for all the
planet-hosting and non-planet-hosting stars in the
Anglo-Australian Planet Search (AAPS) data set with viable
template spectra, so that robust conclusions can be reached about
the differences in their elemental abundances. The inclusion of
these heavy elements begins to extend the spectroscopic studies of
known host stars beyond the iron peak, thus continuing the search
for additional chemical anomalies within these systems, while also
providing an independent check of nucelosynthesis models and
previously published abundances and trends for known host stars.
The lighter elements selected for study (O, Mg and Cr) are
included so as to complement the previous study of \citet{bond}
while also providing valuable information as to the cause of the
metal-enrichment commonly seen in known host stars.

\section{Data}
\subsection{Target Stars}
In this study, we concentrate on the F- and G-type stars, observed
at the 3.9m Anglo-Australian Telescope (AAT) since January 1998 as
part of the AAPS program
\citep{b4,b5,tin1,tin3,tin2,tin05,tin06,j2,j3,j06,c4,mc,jen06,
w07,ot07}. As at January 2007, 31 stars present within the AAPS
sample were known to be planet hosts, of which 28 had spectra
useful for the purposes of this study. Stars known to be young
(age $<$ 3\,Gyr), active (logR\arcmin$_{HK}$$>$ -4.5) or with
other stars within 5\arcsec\ are rejected from the AAPS search.
For a more detailed description of the data and details of the
criteria applied to the AAPS target stars, the reader is referred
to \citet{but96} and \citet{tin05}. Of the 28 host stars
considered here, 26 have had some common elemental abundances
previously determined by other authors.

\subsection{Spectroscopic Analysis}
The method utilised in this study closely follows that outlined in
\citet{bond} who studied Fe, C, Na, Al, Si, Ca, Ti and Ni. Spectra
encompassing the entire visible spectrum from 4820 to 8420\AA\
with a signal-to-noise ratio (S/N) between 200 and 300 per
spectral pixel at resolution $\lambda/\Delta\lambda\approx80000$
were obtained via the University College London Echelle
Spectrograph (UCLES) using the 31.6 line/mm echelle grating as
part of the AAPS program. The raw data were reduced and processed
so as to produce a one-dimensional spectra, suitable for spectral
analysis.

We do differ slightly from \citet{bond} in the method used to
determine the equivalent widths of absorption lines. In
\citet{bond}, we obtained equivalent width estimates via direct
integration over the line. In this study we follow the method of
other similar studies (eg. \citealt{sb,g6,s1}) and make Gaussian
line fits to the spectra using the IRAF task \texttt{splot} in the
package \texttt{noao.onedspec} (due to difficulties in automating
the previous script). This slight difference in methodology does
not produce any significant difference in the final abundances
obtained.

Five heavy elements were analysed for the first time as part of
this study and they were primarily selected based on their process
of production. Y, Zr and Ba are all produced primarily by the
s-process while Eu is primarily produced by the r-process and Nd
is an almost even mix between the two based on Solar System
abundances (\citealt{ar,sim}). The line list utilised in this
study is shown in Table \ref{lines} and is derived from
\citet{gilli} (for Mg and Cr), \citet{red03}(for Y, Zr, Ba, Eu and
Nd) and \citet{den} (for Nd). Additional Ba lines have been used
in other studies, but were neglected here as they gave
consistently lower abundances by approximately 0.60 dex. While the
use of more lines in determining an abundance should produce a
more robust value, we could not determine the cause of this offset
and elected instead to use the single stronger line at 6496.91\AA\
as this line produced a Ba abundance close to the reported solar
value for a solar spectrum. Atomic parameters for each line were
obtained from the NIST Atomic Spectra Database, Version 3.0 (for
O, Mg and Cr), \citet{pits} and \citet{han} (for Y), \citet{red03}
(for Ba), \citet{den} (for Nd) and the Kurucz atomic line
database\footnote{http://www.pmp.uni-hannover.de/cgi-bin/ssi/test/kurucz/sekur.html}
(for Zr and Eu). All of the atomic parameters applied here have
also been utilised in previous studies and produce solar elemental
abundances well within errors of those published elsewhere, when
applied to a solar spectrum, thus giving us confidence in applying
them here.

Elemental abundances were obtained via standard local
thermodynamic equilibrium analysis, as has been done by previous
studies (see \citealt{sb,g6,s1,bond}). A revised version of
Sneden's (1973) MOOG abundance code entitled \texttt{width6}
\citep{r1} was once again used in conjunction with a grid of
\citet{kr} ATLAS9 atmospheres\footnote{http://kurucz.harvard.edu
 or http://www.stsci.edu/hst/observatory/cdbs/k93models.html}
 to obtain the final elemental abundances. As all of the non-host,
and all but 7 of the host stars, had been the subject of an
earlier study (\citealt{bond}), previously published stellar
atmospheric parameters were used. For those stars without
previously determined values, we followed the same method as used
in \citet{bond} to obtain the values and refer the reader to the
paper for more details. As the OI triplet lines are known to
suffer from non-LTE effects, the corrections of \citet{tak} for
$\xi_t$=1km/s and log g=4.0 cm/s$^{2}$ were applied to obtain our
final O abundances. When applied to a solar spectrum, this method
produced abundances in agreement with those of \citet{asp}.

\section{Results}
The stellar elemental abundances are shown in Table \ref{allab}
(in the standard astronomical logarithmic form) and Table
\ref{allsi} (in the more cosmochemically useful form with
abundances normalized to 10$^6$ Si atoms). The notation of $-$ for
an abundance indicates that a value could not be obtained from the
spectrum due to noise. Additionally, for ease of comparison in
Section 5.1, C and Si stellar elemental abundances, along with the
normalized C abundances, for all target stars are presented in
Table \ref{oldab}. The C and Si abundances were previously
published in \citet{bond}.

Other authors have previously determined abundance values for
several of the elements also studied here with many of these
abundances differing from our values. Differences between our
present study and that of others is not a significant issue for
the primary thrust of this paper, which is to compare host and
non-host stellar abundances which have been measured in an
identical fashion. However, in the interests of completeness we
note that Mg abundances were determined for 29 common stars (20
hosts) by \citet{be}, Cr in 28 stars (20 hosts) and Mg in 29 stars
(20 hosts) by \citet{gilli}, Cr in 25 stars (16 hosts) by
\citet{bo}, O in 8 stars (4 hosts) by \citet{ec06}, O and Eu in 4
host stars, Cr and Mg in 1 host star by \citet{gonz} and O in 1
host star by \citet{sb}. The mean differences between our values
and those previously published are shown in Table \ref{difference}
(for those samples having more than one common star) with the
difference being defined as the abundance from this study minus
the published abundance. Generally, our results can be seen to be
in agreement with those previously published for Cr and O with a
significantly larger mean difference occurring for Mg and and a
large deviation occurring for Eu. The differences between our
abundances and those previously published are believed to be due
to the use of a smaller number of lines in determining the
abundance (for Mg), the use of different methods (for O and Eu),
the use of different atomic parameters (for Eu) and the use of
different non-LTE corrections (for O).

\section{Host Star Enrichment}
\subsection{Enrichment over Solar}
The mean and median abundances, standard deviation and the
difference between the host and non-host stars for all of our
target stars can be seen in Table \ref{stat} for each element. The
quoted uncertainties are the standard error in the mean, and the
median uncertainty from the algorithm of \citet{kd}\footnote{For a
distribution with N values, the error in the median is the range
in values on either side of the median which contains ($\surd$N)/2
values }. The mean values of this study for the known host stars
are all consistent to within the 1$\sigma$ value of those listed
by \citet{be} and \citet{gilli}. The data show that in general
known extrasolar planetary host stars differ only slightly from
the mean solar abundance patterns with all of the median
abundances being well within 1$\sigma$ of the solar abundance - as
concluded by previous studies (eg. \citealt{ec,bo}). This is
reassuring as the Sun is itself is obviously a planetary host star
with abundances enhanced over those of most other stars in the
solar neighbourhood (based on the abundances of \citealt{asp}). In
many respects, therefore, the Sun is \emph{not} a typical field
star, based on its abundances and its multiple planetary
companions. The largest enrichment over solar is seen in Nd and
Zr, with Cr showing a smaller enrichment and Mg showing minimal
enrichment. Eu showed the largest depletion relative to solar
values, with Y and Ba also showing mild to moderate depletions. Only O produced a mean abundance equal to the Solar abundance.

Similarly, the non-host stars can also be seen be depleted when
compared to solar abundances for almost all of the elements
studied, with the largest depletion being $-$0.16 for Y and Eu. It is
also worth noting that three of the five heavy elements examined
show a mean depletion relative to solar for both the host and
non-host stars. Of these three, two are produced by the s-process
(Y \& Ba) with the remaining element (Eu) produced by the
r-process.

\subsection{Enrichment over Non-Host Stars}
A more powerful comparison is obtained by comparing our host and
non-host populations to each other. On doing so, it can be clearly
seen that host stars are systematically enriched over non-host
stars in all elements studied. The enrichment ranges in size from
0.06 (for O) to 0.11 for (for Cr and Y) (see Table \ref{stat}).

This difference between the host and non-host populations can also
be seen in the results of the Kolmogorov-Smirnov (K-S) statistical
test. Designed to test whether two populations were drawn from the
same parent sample, the K-S test showed a significant difference
between the host and non-host samples with probabilities of a
different parent sample ranging from 91.2\% for O to 99.99\% for Y
(see Table \ref{stat}). This supports our claim that host star
elemental abundances are significantly different to those of
non-host stars and furthermore that host stars are enriched over
non-host stars for all elements studied.

\section{Elemental Trends}
\begin{figure}
\plotone{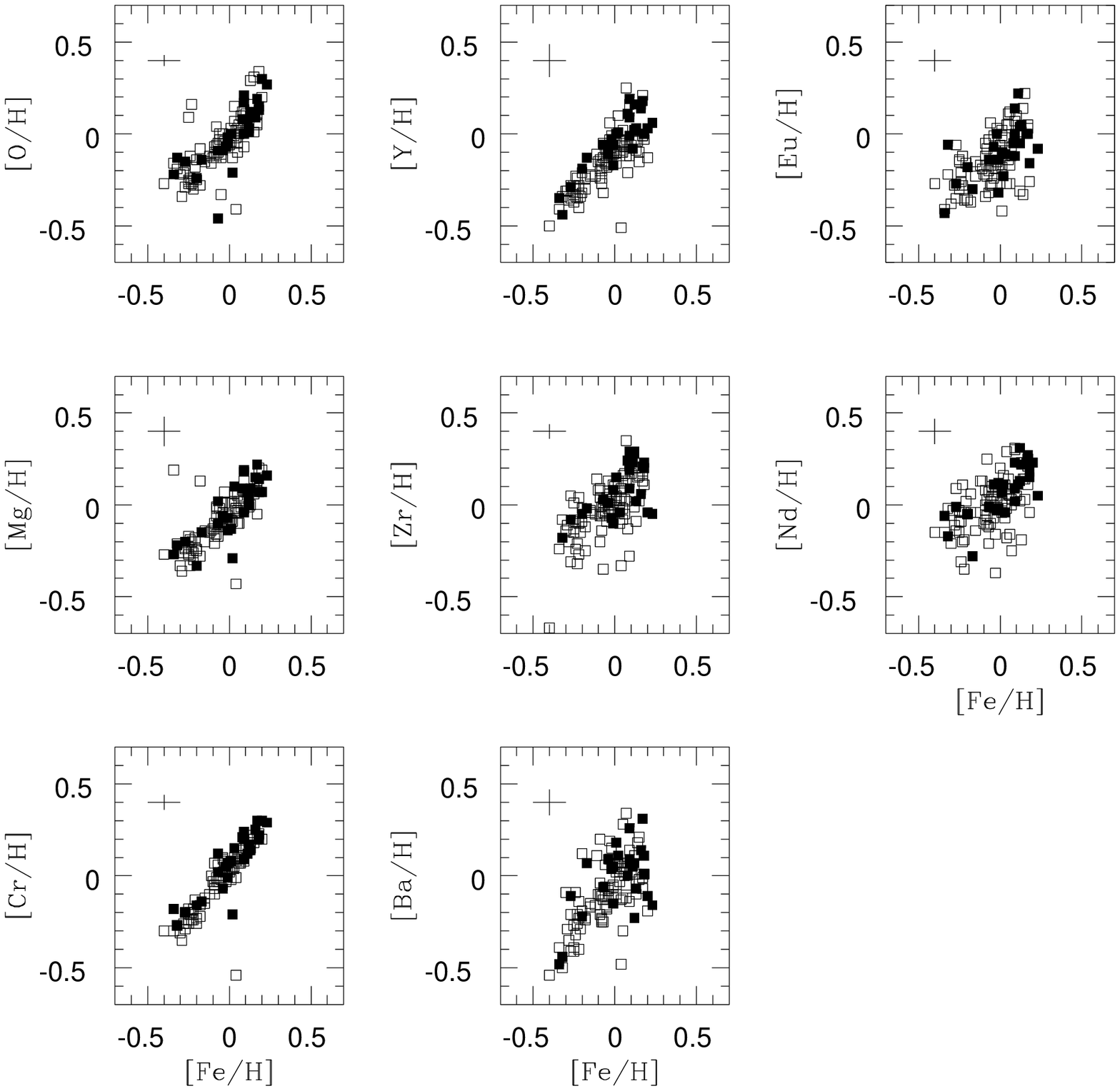} \caption{[X/H] vs. [Fe/H] plots for all elements
studied. Open squares represent non-host stars and filled squares
represent host stars. Typical error bars are shown in the upper
left of each panel. \emph{Left Column:} O, Mg and Cr. \emph{Center
Column:} Y, Zr and Ba. \emph{Right Column:} Eu and Nd.}
\label{allh}
\end{figure}

\begin{figure}
\plotone{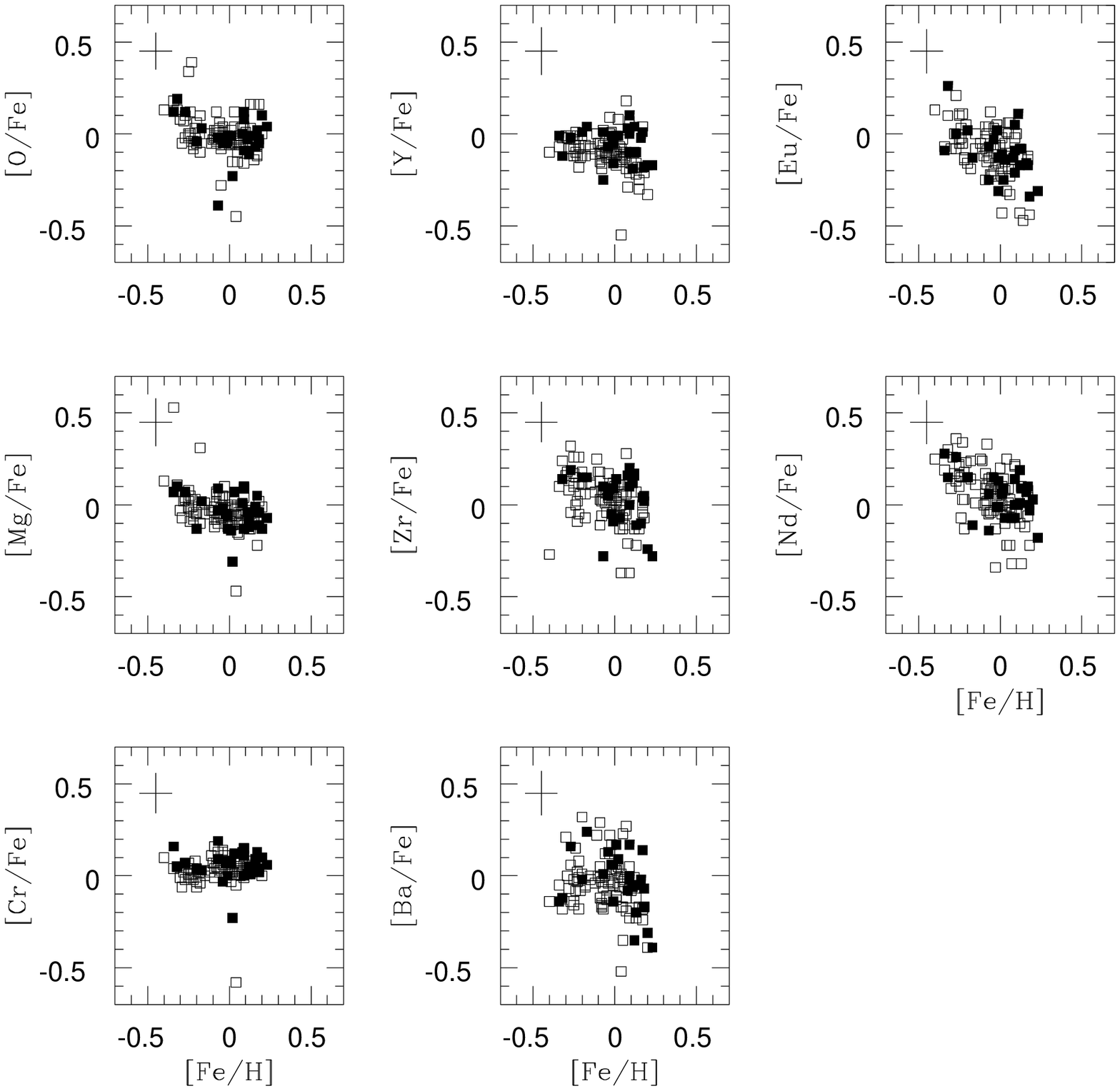} \caption{[X/Fe] vs. [Fe/H] plots for all elements
studied. Open squares represent non-host stars and filled squares
represent host stars. Typical error bars are shown in the upper
left of each panel. \emph{Left Column:} O, Mg and Cr. \emph{Center
Column:} Y, Zr and Ba. \emph{Right Column:} Eu and Nd.}
\label{allfe}
\end{figure}

Plots of our results are shown in Figures \ref{allh} and
\ref{allfe} - in Figure \ref{allh} we present [X/H] versus [Fe/H]
as more commonly used in previous studies of planet host star
abundances, while in Figure \ref{allfe} we present [X/Fe] versus
[Fe/H] as usually analysed in cosmochemical studies. Two
significant outliers can be seen - one host and one non-host
sitting below the general trend for O, Cr and Mg. These stars are
HD142415 ([Fe/H]=0.02, host) and HD199288 ([Fe/H]=0.04, non-host).
These stars can be seen to be depleted (compared to solar
abundances) in the majority of elements studied here except for
Fe, suggesting that they have formed from Fe-rich precursor
material based on the assumption that it is easier to enrich one
element than it is to deplete seven elements. The fact that
HD142415 is also mildly enriched in both Ba and Y (with no Zr
abundance available) could also possibly indicate that the
material had been processed through an s-process environment,
either an AGB star or the He-burning stage of a larger mass star.

\emph{Fe and Fe precursors:} The [X/H] trends in Figure \ref{allh}
are in agreement with the understanding we currently have about
the nucelosynthetic origin of the elements. All of the elements
located before the Fe peak (here O, Mg and Cr) increase linearly
with increasing [Fe/H] with Pearson product-moment correlation
coefficients (r) above 0.70 for both host and non-host stars. This
can be easily understood as the stellar evolutionary processes
that serve to increase the amount of Fe present in later
generations of stars also produce the pre-iron peak elements in
various amounts. Thus as the stellar Fe content increases, so too
would the amount of pre-iron peak elements (neglecting any unusual
mixing or other nebula interactions).

\emph{s-process:} The r- and s-process elements, however, are less
certain. All three s-process elements (Y, Zr and Ba) still display
to varying degrees the same trend of increasing abundance with
increasing [Fe/H] as the pre-iron peak elements. One possible
explanation for this observed trend is that the increase in
s-process elemental abundances is due to the increase in the
number of seed nuclei (e.g. Fe atoms) available. Due to the nature
of the s-process, it is reliant on the sufficient availability of
seed nuclei to be able to proceed. Thus as the metallicity
increases, so too does the abundance of s-process species.

\emph{r-process:} Unlike the s-process elements, the r-process
element (Eu) and the mixed source element (Nd) do not display a
strong correlation with increasing [Fe/H]. Observations of metal
poor stars have shown that the abundances of s-process elements
such as Y and Ba decrease faster with metallicity than the
abundances of r-process elements such as Eu (\citealt{spite}).
This has been attributed to a lack of appropriate seed nuclei
inhibiting the s-process significantly more than the r-process. We
are alternatively extending this into the metal-rich regime to
conclude that the r-process is not as reliant on the presence of
elements such as Fe, thus explaining its lack of a strong
dependance upon metallicity.

From Figure \ref{allfe}, it can be seen that the overall [X/Fe]
trends visible here are in good agreement with those identified by
\citet{bo} and \citet{gilli} (which are discussed in more detail
below). They can also be seen as extensions into the high
metallicity region of those trends identified by \citet{red06}.

\subsection{Lighter Element Trends}
In addition to examining the nature of the general trend of
increasing elemental abundances with increasing metallicity, we
can also make some basic determinations about the nature of the
various nucelosynthesis processes occurring within the precursors
to these systems by considering the more subtle, second order
trends present within the data.

\textit{O:} Based on our data, [O/Fe] displays a weakly correlated
decreasing trend with increasing [Fe/H] for both the host stars
(slope=$-$0.16, r=$-$0.22) and non-host stars (slope=$-$0.24,
r=$-$0.32). Previous studies have hinted at the possibility of a
plateau starting at [Fe/H]$\sim$0
(\citealt{red03},\citealt{red06}). However, the overlap between
our sample and those previously published is not large enough to
allow us to undertake any meaningful comparison. It is also worth
noting that while the solar C/O ratio is 0.54 (\citealt{asp}),
using the [C/H] values previously published by \citet{bond} (see
Table \ref{oldab}), the C/O values of the host stars studied here
range from 0.40 to 0.89. This variation has the potential to
greatly impact the C chemistry of the proto-stellar disc and thus
also any terrestrial planets forming in the system. Further
studies into this subject are currently underway (see
\citealt{bond2}).

\textit{Mg:} [Mg/Fe] can be seen to display a weakly correlated
trend of decreasing with increasing [Fe/H] values producing r
values of $-$0.32 (host stars) and $-$0.44 (non-host stars). The
distribution observed for Mg is in good agreement (for the [Fe/H]
regions in common) with those of previous studies who observed a
decrease in [Mg/Fe] with increasing [Fe/H] up to
[Fe/H]$\sim$$-$0.10 before both distributions plateaued
(\citealt{be,gilli,red06}).

Also of interest is the Mg/Si ratio of our target stars as it has
the potential to greatly affect the chemical evolution of any
terrestrial planets forming in the system. A high Mg/Si ratio will
result in all of the available Si forming olivine
((Fe,Mg)$_{2}$SiO$_{4}$) with excess Mg still being available,
while a low Mg/Si ratio will result in all of the available Mg
forming enstatite (MgSiO$_{3}$) with excess Si forming SiO$_{2}$.
Utilising the Si abundances previously published in \citet{bond}
(see Table \ref{oldab}), the host stars of this study were found
to have Mg/Si ratios ranging from 0.46 to 1.26, resulting in
potentially large variations in the nature of any terrestrial
planets forming in these systems. A separate study examining this
issue is currently underway (see \citealt{bond2}).

\textit{Fe Group (Cr):} From Figure \ref{allfe}, it can be seen
that (with the exception of one outlier) [Cr/Fe] displays no
significant trend, remaining largely unchanged over all of the
[Fe/H] values considered here. This is in good agreement with
\citet{red03,ben} and \citet{gilli}. The lack of a statistically
significant trend with [Fe/H] is confirmed by the low r value of
$-$0.04 for hosts and 0.00 for non-hosts.

\subsection{Heavy Element Trends}

From Figure \ref{allfe}, it can be seen that all of the heavy
elements display varying degrees of a weakly negative to
non-existent correlation with [Fe/H]. The r values produced range
from $-$0.22 (Y and Ba) to $-$0.56 (Nd) indicating that the
correlation is not strong.

These trends are in agreement with \citet{ap} and \citet{ben} over
the range of metallicity values in common with this study. We do
differ slightly from some previous studies in that we observed a
stronger decrease in [Y/Fe] with increasing [Fe/H] than was
observed by \citet{ben} and disagree with previous studies who
observed no significant trend with [Fe/H] for both [Ba/Fe] and
[Nd/Fe] (\citealt{red03}). We do observe trends in both of these
samples and the difference is attributed to the fact that we are
examining a different metallicity region to that of \citet{red03}.
Our population is largely concentrated in the region of
[Fe/H]$>$0, while the sample of \citet{red03} almost exclusively
has values of [Fe/H]$<$0. However, it is worth restating that the
general trends in the neutron capture elements identified here are
the same as those previously identified for other solar-type
stars. This implies that the host stars examined in this paper
follow the same trends as other field stars but with a bias
towards the high metallicity region. Additionally, the high degree
of scatter observed in these samples is also in agreement with
previous studies.

\begin{figure}
\plotone{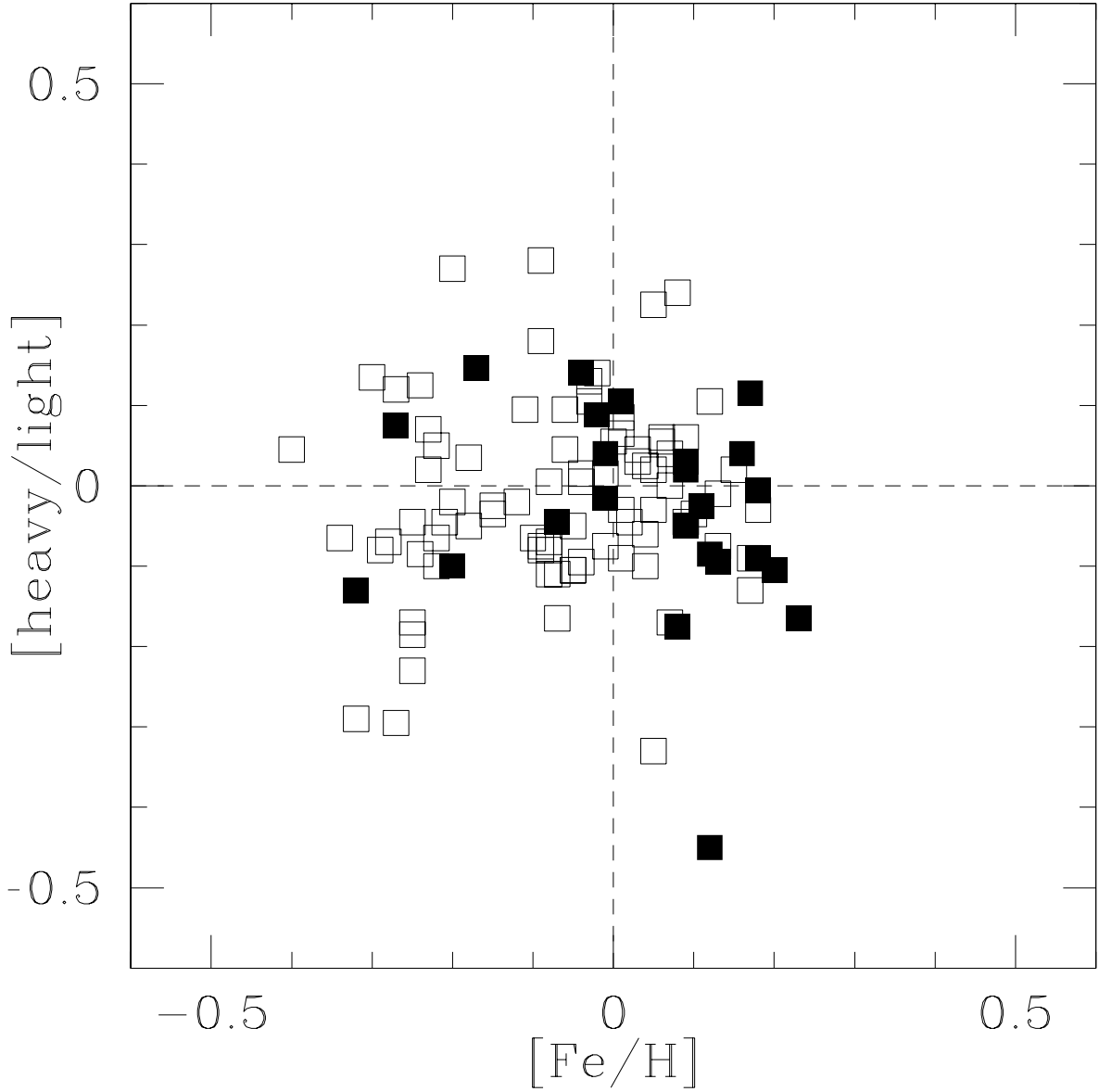} \caption{[heavy/light] vs. [Fe/H] plots for the
five heavy elements examined as part of this study. For the
definition of [heavy/light], please see the text. Open squares
indicate non-host stars, filled squares indicate host stars and
dashed lines indicate solar values.} \label{hl}
\end{figure}

Also of interest is the ratio of the heavy to light s-process
elements as each are thought to be produced in slightly different
stellar settings. The abundance of the heavy s-process element Ba
to the light s-process elements Y and Zr is shown in Figure
\ref{hl} as [heavy/light] where:
\begin{eqnarray}
\nonumber {\rm [heavy/light]} = \rm [Ba/H] - \frac{\rm
[Y/H]+[Zr/H]}{2}
\end{eqnarray}\\

From Figure \ref{hl}, it can be clearly seen that there is no
dependance on metallicity, with values scattering around the solar
value (0.0 by definition). This is in agreement with
\citet{red03}. Thus we support their conclusion that the neutron
exposure in AGB stars is independent of the metallicity of the
star itself, assuming (as \citealt{red03} did) that AGB stars are
the primary source of both the heavy and the light s-process
elements.

\subsection{Correlation with Planetary Parameters}
\begin{figure}
\plotone{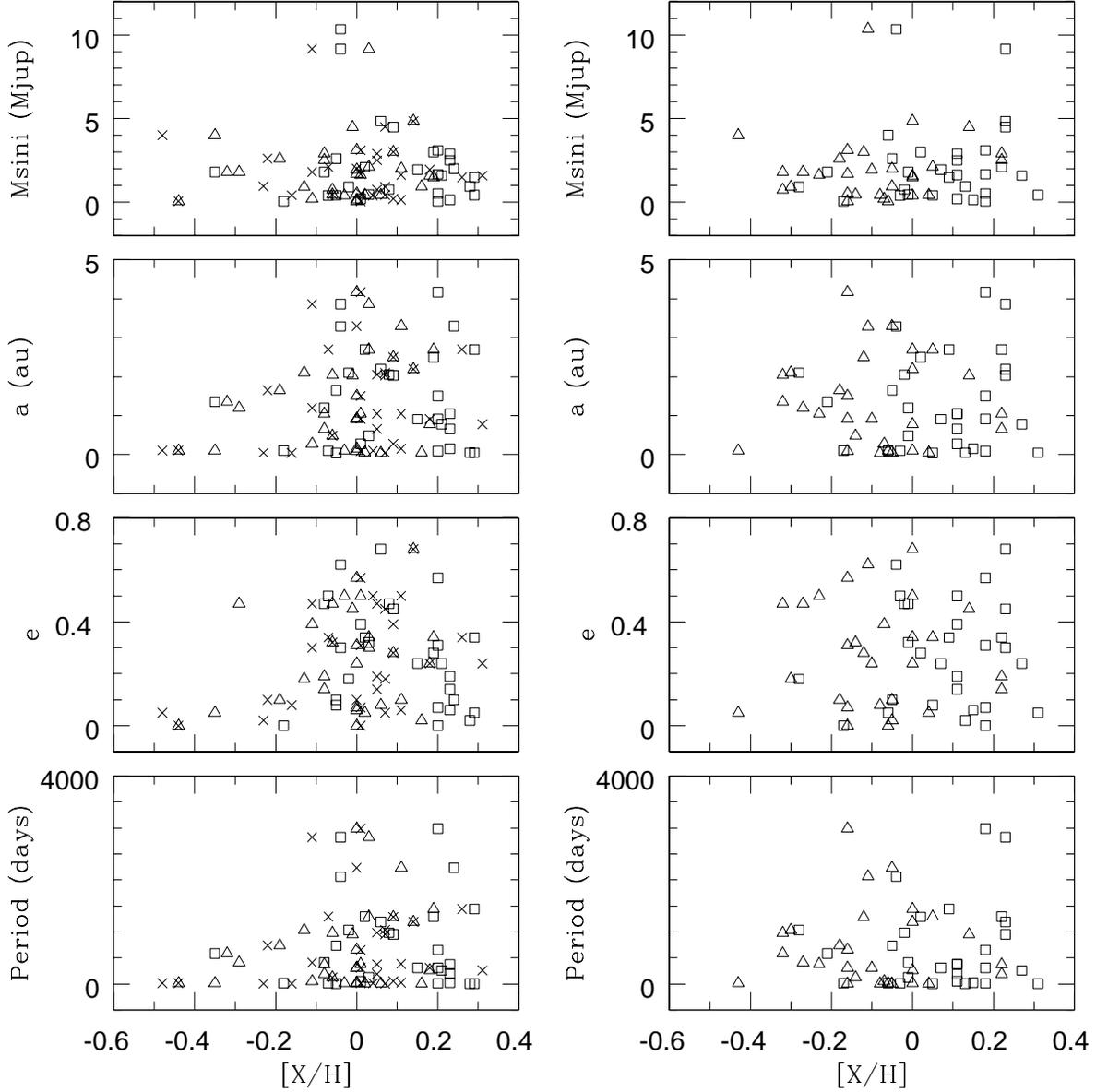} \caption{Orbital properties of extrasolar
planetary systems vs. abundance of the heavy elements. HD164427
has been omitted. {\em Left}: s-process elements Y (triangles), Zr
(squares) and Ba (crosses). {\em Right}: r- and mixed process
elements Eu (triangles) and Nd (squares). {\em From top to
bottom}: Msin{\em i} of the planet, orbital semi-major axis of the
planet, eccentricity of the orbit of the planet and period of the
planet.} \label{orbits}
\end{figure}

Figure \ref{orbits} shows plots of [X/H] against planetary
parameters (M$\sin i$, semi-major axis $a$, eccentricity and
planetary period) for the 5 heavy elements considered in this
study. HD164427 was omitted as its companion is believed to be a
brown dwarf, not a gas giant. As can be seen visually and by the
low r values (all $\leq$0.48 with most $<$0.15), no statistically
significant correlations exist between these abundances and any of
the orbital parameters. This agrees with previous studies of other
elements (eg. \citealt{re,s2,fv}).

\subsection{Correlation with Stellar Parameters}
It is well known that the stellar atmospheric parameters
(specifically Teff and log g) have the potential to drastically
alter photospherically determined stellar abundances. As such, we
examined the abundances presented here and as both samples
produced r$^2$ correlation coefficients less than 0.5, we
concluded that there is no statistically significant trends
present with either the stellar Teff or log g values.

\section{Discussion}
The host stars studied here do not significantly deviate from
previously established galactic chemical evolution trends.
Instead, they can be regarded as being extensions of many of those
trends into metallicities above solar. This lack of deviation from
previously known trends strongly suggests that while they are more
metal enriched than other stars not known to host planets, the
host stars themselves have not systematically undergone any
extraordinary chemical processing during their growth and
evolution (in agreement with \citealt{rob}). The conclusion that
planetary hosts stars are essentially ``normal'' may indeed
suggest that planetary formation is a normal result of the star
formation process. Of course, this does not exclude planet
formation at lower metallicity values (as planets have been
detected orbiting stars with metallicities significantly below
solar) nor does it guarantee planet formation at high metallicity
values.

There are two primary hypotheses that have been offered to explain
the observed high metallicity trend. The first is the
``pollution'' model which posits metal-rich material being added
to the photosphere as a consequence of planetary formation
(\citealt{la,g6,mu}). The second explanation is commonly referred
to as the primordial model and it suggests that the gas cloud from
which these systems formed was metal enriched, resulting in the
star itself being enriched in the same elements (\citealt{s1}).
Our conclusion that these host stars exhibit normal chemical
evolution trends and that they are simply the metal-rich members
of field star population lends support to the primordial model.
One would expect that pollution of the stellar photosphere would
produce deviations from the galactic evolutionary trends. To date
no such trends have been observed. Additionally, we also observe
that the abundances of the more volatile elements (such as O)
increase with increasing metallicity for both the host and
non-host stars. This would not be the case in the pollution model
as it is likely that only the more refractory elements (such as Fe
and Ni) would remain in the solid form as they migrated towards
the star (and thus be deposited in the stellar photosphere) while
the more volatile elements would be evaporated before they could
be incorporated into the stellar photosphere. As such, we would
expect to see enrichment only in the refractory elements and not
the volatile elements if the pollution model is accurate.
Furthermore, the fact that we observe no trends with metallicity
in the orbital parameters of the remaining planets is difficult
for the pollution model to explain. It is hard to imagine a
situation whereby almost \textit{all} planetary host stars
accreted a significant amount of material during planetary
formation without affecting the orbital parameters of the
remaining planets. For these reasons, we agree with previous
studies (such as \citealt{s1,s2,s04,s05} and \citealt{fv}) and
support the primordial model for explaining the metal enrichment
in planetary host stars.

The abundances reported here also impact on terrestrial planet
formation and evolution within these systems. Those with low Mg/Si
ratios will have terrestrial planets dominated by enstatite (with
a small amount of Mg-rich olivine also present) with other
Si-based species also available (a composition similar to the
Earth's crust), while those with high Mg/Si ratios will have
olivine-dominated planets with other Mg-rich species also present
(a composition similar to the Earth's mantle). Similarly, a high
C/O ratio will result in planets with greatly increased C contents
due to solid C being incorporated into the planet itself. While
the detailed consequences of these examples have not yet been
fully examined, it is conceivable that any terrestrial planets
forming in these systems could differ from currently known
terrestrial planets in terms of their rheology (thus possibly
affecting the tectonics of such a planet) and the nature of
volcanic activity, based on the varying silica contents of the
magma. The full implications of such a chemical composition for
the evolution of the terrestrial planets themselves is the subject
of ongoing research.

\section{Summary}
In this paper, we have presented elemental abundances for 8
elements, including 5 heavy elements produced by the r- and
s-processes, in 28 planetary host stars and 90 non-host stars from
the AAPS. We conclude that while the elemental abundances of the
planetary host stars are only slightly different from solar
values, the host stars are enriched over the non-hosts stars in
all elements studied  with the mean difference varying from 0.06
dex to 0.11 dex.

Additionally, we also considered the trends of the abundances
(both [X/H] and [X/Fe]) with [Fe/H] and found these to be largely
in keeping with known galactic chemical evolution trends. This
implies that these systems have followed normal evolutionary
pathways and are not significantly or unusually altered. This
leads us to conclude that not only are the abundance trends we are
observing primordial in origin and represent the initial
composition of the gas nebula that produced the star and its
planets but that planetary formation may also be a natural
companion to the evolution of stellar material.

\acknowledgments

We would like to thank Sean Ryan for his assistance with this
study by providing the stellar atmosphere and abundance code,
along with incredibly patient help. We would also like to thank
the anonymous reviewer for their useful suggestions and advice.
The Anglo-Australian Planet Search team would like to acknowledge
support by the partners of the Anglo-Australian Telescope
Agreement (CGT and HRAJ); NASA grant NAG5-8299, NSF grant AST 95-
20443 (GWM), NSF grant AST 99-88087 (RPB), PPARC grant
PP/C000552/1 (HRAJ, CGT, SJO) and ARC grant DP0774000 (CGT). This
research has made use of the SIMBAD database, operated at CDS,
Strasbourg, France, and the NASA Astrophysics Data System.

\clearpage

\begin{deluxetable}{cccccc}
\tablewidth{0pt} \tablecaption{Line list used for chemical
abundance analysis\label{lines}} \tablehead{ \colhead{$\lambda$} &
\colhead{Log {\it gf}}      & \colhead{$\chi_{\it l}$} &
\colhead{$\lambda$}  & \colhead{Log {\it gf}} &
\colhead{$\chi_{\it l}$} \\
\colhead{(\AA)}  &  & \colhead{(eV)} & \colhead{(\AA)} &  &
\colhead{(eV)}} \startdata
\multicolumn{1}{l}{\textbf{Mg\,{\sc I}:}} & &&   \multicolumn{1}{l}{\textbf{Y\,{\sc II}:}}  \\
5711.09 & -1.71 & 4.35  & 4854.87 &-0.01& 0.99 \\
6318.72 & -1.99 & 5.11 &4900.12 &-0.09& 1.03 \\
\multicolumn{1}{l}{\textbf{Cr\,{\sc I}:}}  & && 5087.43 &-0.17& 1.08 \\
5304.18 & -0.69 & 3.46 &5200.42 &-0.49& 0.99 \\
5312.87 & -0.56 & 3.45 &5402.78 &-0.63& 1.84 \\
5318.81 & -0.69 & 3.44 &\multicolumn{1}{l}{\textbf{Zr\,{\sc II}:}}  \\
5783.09 & -0.50 & 3.32 &5112.28 &-0.59& 1.66 \\
5783.89 & -0.29 & 3.32&   \multicolumn{1}{l}{\textbf{Eu\,{\sc II}:}}  \\
\multicolumn{1}{l}{\textbf{O\,{\sc I}:}}   & && 6645.13 &0.20& 1.38 \\
7771.94 & 0.37 & 9.15 & \multicolumn{1}{l}{\textbf{Nd\,{\sc II}:}}  \\
7774.17 & 0.22 & 9.15 &4914.18 &-0.70& 0.38 \\
7775.39 & 0.002 & 9.15 &5234.19 &-0.51& 0.55 \\
\multicolumn{1}{l}{\textbf{Ba\,{\sc II}:}} & \\
6496.91 &-0.41& 0.60 \\
\tableline
\enddata
\end{deluxetable}

\clearpage
\begin{deluxetable}{rrrrrrrrrrr}
\tabletypesize{\scriptsize} \rotate \tablewidth{0pt}
\tablecaption{Stellar Abundances for all Target Stars. A $-$
indicates that a value could not be obtained from the
spectrum.\label{allab}} \tablehead{ \colhead{HD} & \colhead{[O/H]}
& \colhead{[Cr/H]}          & \colhead{[Mg/H]}  & \colhead{[Ba/H]}
& \colhead{[Y/H]}    & \colhead{[Zr/H]}  & \colhead{[Eu/H]}  &
\colhead{[Nd/H]}} \startdata
\multicolumn{2}{l}{Host Stars:}\\
2039    &   0.09    $\pm$   0.04    &   0.25    $\pm$   0.07    &   0.15    $\pm$   0.07    &   0.14    $\pm$   0.03    &   0.14    $\pm$   0.09    &   0.06    $\pm$   0.03    &   0.00    $\pm$   0.02    &   0.23    $\pm$   0.25  \\
4308    &    $-$0.13    $\pm$   0.02    &    $-$0.27    $\pm$   0.02    &    $-$0.22    $\pm$   0.09    &    $-$0.44    $\pm$   0.02    &    $-$0.44    $\pm$   0.08    &    $-$0.18    $\pm$   0.03    &    $-$0.06    $\pm$   0.01    &    $-$0.17    $\pm$   0.16     \\
10647   &    $-$0.14    $\pm$   0.02    &    $-$0.14    $\pm$   0.06    &    $-$0.15    $\pm$   0.06    &   0.07    $\pm$   0.03    &    $-$0.13    $\pm$   0.12    &    $-$0.02    $\pm$   0.04    &    $-$0.30    $\pm$   0.01    &    $-$0.28    $\pm$   0.20     \\
13445   &    $-$0.22    $\pm$   0.03    &    $-$0.18    $\pm$   0.03    &    $-$0.27    $\pm$   0.09    &    $-$0.48    $\pm$   0.08    &    $-$0.35    $\pm$   0.10    &      \multicolumn{1}{c}{$-$}          &    $-$0.43    $\pm$   0.06    &    $-$0.06    $\pm$   0.03     \\
17051   &   0.00    $\pm$   0.01    &   0.08    $\pm$   0.06    &    $-$0.13    $\pm$   0.01    &   0.18    $\pm$   0.08    &   0.00    $\pm$   0.12    &   0.15    $\pm$   0.04    &    $-$0.10    $\pm$   0.03    &   0.07    $\pm$   0.10     \\
23079   &    $-$0.24    $\pm$   0.02    &    $-$0.16    $\pm$   0.05    &    $-$0.33    $\pm$   0.01    &    $-$0.22    $\pm$   0.03    &    $-$0.19    $\pm$   0.10    &    $-$0.05    $\pm$   0.09    &    $-$0.18    $\pm$   0.04    &    $-$0.05    $\pm$   0.10     \\
20782   &    $-$0.46    $\pm$   0.04    &   0.12    $\pm$   0.05    &   0.02    $\pm$   0.07    &    \multicolumn{1}{c}{$-$}            &    $-$0.32    $\pm$   0.08    &    $-$0.35    $\pm$   0.05    &    $-$0.32    $\pm$   0.06    &    $-$0.21    $\pm$   0.05     \\
\enddata
\end{deluxetable}

\clearpage
\begin{deluxetable}{rrrrrrrrrrr}
\tabletypesize{\scriptsize} \rotate \tablewidth{0pt}
\tablecaption{Stellar Abundances normalized to 10$^6$ Si atoms.
Solar values are based on the abundances in
\citet{asp}\label{allsi}} \tablehead{ \colhead{HD} & \colhead{Fe}
& \colhead{O} & \colhead{Mg}          & \colhead{Cr} & \colhead{Y}
& \colhead{Zr}    & \colhead{Ba}  & \colhead{Nd}  & \colhead{Eu}}
\startdata
\multicolumn{2}{l}{Host Stars:}\\
Solar   &   8.71    x10$^   5   $&  1.41    x10$^   7   $&  1.05    x10$^   6   $&  1.35    x10$^   4   $&  5.01    &   12.02   &   4.57    &   0.87    &   0.10     \\
2039    &   1.00    x10$^   6   $&  9.12    x10$^   6   $&  7.76    x10$^   5   $&  1.26    x10$^   4   $&  3.63    &   7.24    &   3.31    &   0.78    &   0.05     \\
4308    &   8.32    x10$^   5   $&  1.48    x10$^   7   $&  8.91    x10$^   5   $&  1.02    x10$^   4   $&  2.57    &   11.22   &   2.34    &   0.83    &   0.13     \\
10647   &   9.77    x10$^   5   $&  1.70    x10$^   7   $&  1.23    x10$^   6   $&  1.62    x10$^   4   $&  6.17    &   18.88   &   8.91    &   0.76    &   0.09     \\
13445   &   8.32    x10$^   5   $&  1.78    x10$^   7   $&  1.17    x10$^   6   $&  1.86    x10$^   4   $&  4.68    &    \multicolumn{1}{c}{$-$}    &   3.16    &   1.57    &   0.08     \\
17051   &   7.94    x10$^   5   $&  1.15    x10$^   7   $&  6.31    x10$^   5   $&  1.32    x10$^   4   $&  4.07    &   13.80   &   5.62    &   0.83    &   0.07     \\
23079   &   1.07    x10$^   6   $&  9.77    x10$^   6   $&  5.89    x10$^   5   $&  1.12    x10$^   4   $&  3.89    &   12.88   &   3.31    &   0.93    &   0.08     \\
\enddata
\end{deluxetable}

\clearpage
\begin{deluxetable}{rrrrrrrr}
\tabletypesize{\scriptsize} \tablewidth{0pt} \tablecaption{C and
Si Abundances from Bond et al. (2006).\label{oldab}} \tablehead{
\colhead{HD} & \colhead{[C/H]} & \colhead{C atoms} &
\colhead{[Si/H]} & \colhead{HD} & \colhead{[C/H]} & \colhead{C
atoms} & \colhead{[Si/H]} } \startdata
\multicolumn{2}{l}{Host Stars:}&&&\multicolumn{2}{l}{Non-Host Stars (cont.):} \\
Solar   &               &   7.59    x10$^   6   $&              &   31827   &   0.37    $\pm$   0.12    &   5.37    x10$^   6   $&  0.52    $\pm$   0.14     \\
2039    &   0.22    $\pm$   0.16    &   6.61    x10$^   6   $&  0.28    $\pm$   0.12    &   33811   &   0.03    $\pm$   0.11    &   4.07    x10$^   6   $&  0.30    $\pm$   0.06     \\
4308    &   $-$0.16 $\pm$   0.08    &   7.41    x10$^   6   $&  $-$0.15 $\pm$   0.06    &   36108   &   $-$0.13 $\pm$   0.06    &   8.71    x10$^   6   $&  $-$0.19 $\pm$   0.05     \\
10647   &   $-$0.14 $\pm$   0.06    &   9.12    x10$^   6   $&  $-$0.22 $\pm$   0.09    &   38283   &   $-$0.07 $\pm$   0.03    &   9.12    x10$^   6   $&  $-$0.15 $\pm$   0.03     \\
13445   &   $-$0.17 $\pm$   0.10    &   1.07    x10$^   7   $&  $-$0.32 $\pm$   0.10    &   38382   &   0.03    $\pm$   0.05    &   8.32    x10$^   6   $&  $-$0.01 $\pm$   0.04     \\
17051   &   0.09    $\pm$   0.11    &   7.59    x10$^   6   $&  0.09    $\pm$   0.07    &   38973   &   $-$0.01 $\pm$   0.06    &   7.24    x10$^   6   $&  0.01    $\pm$   0.09     \\
23079   &   $-$0.07 $\pm$   0.10    &   7.76    x10$^   6   $&  $-$0.08 $\pm$   0.05    &   42902   &   0.30    $\pm$   0.13    &   7.24    x10$^   6   $&  0.32    $\pm$   0.08     \\
\enddata
\end{deluxetable}

\clearpage

\begin{deluxetable}{ccrc}
\tablewidth{0pt} \tablecaption{Mean difference from published
abundances\label{difference}} \tablehead{ \colhead{Study} &
\colhead{Element}      & \colhead{Difference} & \colhead{Sample
Size} } \startdata
Beri\~{a}o et al. (2005) & Mg & $-$0.18$\pm$ 0.02&29\\
Gilli et al. (2006) & Cr & 0.00$\pm$ 0.02&28\\
& Mg&$-$0.18$\pm$ 0.02&29\\
Bodaghee et al (2003)&Cr&0.02$\pm$ 0.01&25\\
Ecuvillon et al (2006)&O&0.11$\pm$ 0.07&8\\
Gonzalez \& Laws (2007)&O&$-$0.03$\pm$ 0.04&4\\
&Eu&$-$0.22$\pm$ 0.16&4\\
\tableline
  \enddata
\end{deluxetable}

\clearpage

\begin{deluxetable}{lllcclllcc}
\tabletypesize{\scriptsize} \rotate \tablewidth{0pt}
\tablecaption{Statistical analysis of abundance distributions.
Difference is defined as host star abundance $-$ non-host star
abundance. Numbers in parentheses indicate the sample size. K$-$S
test values are the confidence percentages that the two samples do
\textit{not} come from the same parent population.\label{stat}}
\tablehead{ &\colhead{Non-Planetary} & \colhead{Planetary Hosts} &
\colhead{Difference} & \colhead{K$-$S} && \colhead{Non-Planetary}
& \colhead{Planetary
Hosts}      & \colhead{Difference} & \colhead{K$-$S}\\
 & \colhead{Hosts}& && \colhead{Result} & & \colhead{Hosts}& && \colhead{Result} }
\startdata
O:&&&&&Zr:\\
Mean   & $-$0.06$\pm$ 0.02 (90)& $+$0.00$\pm$ 0.03 (27)& 0.06 &
91.2&Mean   & $-$0.03$\pm$ 0.02 (85)& $+$0.06$\pm$ 0.03 (26)& 0.09&97.5\\
Median & $-$0.05$\pm$ 0.02     & $+$0.00$^{+0.04} _{-0.03}$   &
0.05& &Median & $-$0.01$^{+0.03} _{-0.02}$ & $+$0.05$^{+0.03} _{-0.06}$  & 0.06&     \\
Std. Dev. & \multicolumn{1}{c}{0.15} & \multicolumn{1}{c}{0.17} &
& & Std. Dev. & \multicolumn{1}{c}{0.15} & \multicolumn{1}{c}{0.16} & &\\ \\
Mg:&&&&&Ba:\\
Mean   & $-$0.09$\pm$ 0.01 (90)& $-$0.02$\pm$ 0.03 (28)& 0.07 & 96.82&Mean   & $-$0.11$\pm$ 0.02 (85)& $-$0.01$\pm$ 0.04 (26)& 0.10 & 98.2\\
Median & $-$0.10$^{+0.04} _{-0.01}$     & $+$0.01$^{+0.06}
_{-0.07}$ & 0.11 &   &Median   & $-$0.11$\pm$ 0.02     & $+$0.05$^{+0.04} _{-0.05}$ & 0.16&       \\
\enddata
\end{deluxetable}

\clearpage

\clearpage

\clearpage

\clearpage

\end{document}